\documentclass[runningheads]{llncs}
\usepackage{graphicx}
\usepackage{multirow}
\usepackage{verbatim}
\usepackage{cite}
\usepackage{amssymb}
\graphicspath{{figures/}}
\usepackage{url}
\usepackage{booktabs} % For the tables

\begin{document}

\title{Analysis of Hand-Crafted and Automatic-Learned Features for Glaucoma Detection Through Raw Circumpapillary OCT Images}
\titlerunning{Hybrid learning for glaucoma detection from OCT images}

\author{Gabriel Garc\'{i}a\inst{1}, Adri\'{a}n Colomer\inst{1}, Valery Naranjo\inst{1}}
\authorrunning{Gabriel Garc\'{i}a \textit{et al.}}
\institute{$^{1}$Instituto de Investigación e Innovación en Bioingeniería (I3B), Universitat Politècnica de València, Camino de Vera s/n, 46022, Valencia, Spain.\\
\email{jogarpa7@i3b.upv.es}}

\maketitle              % typeset the header of the contribution

\begin{abstract}
Taking into account that glaucoma is the leading cause of blindness worldwide, we propose in this paper three different learning methodologies for glaucoma detection in order to elucidate that traditional machine-learning techniques could outperform deep-learning algorithms, especially when the image data set is small. The experiments were performed on a private database composed of 194 glaucomatous and 198 normal B-scans diagnosed by expert ophthalmologists. As a novelty, we only considered raw circumpapillary OCT images to build the predictive models, without using other expensive tests such as visual field and intraocular pressure measures. The results ratify that the proposed hand-driven learning model, based on novel descriptors, outperforms the automatic learning. Additionally, the hybrid approach consisting of a combination of both strategies reports the best performance, with an area under the ROC curve of 0.85 and an accuracy of 0.82 during the prediction stage.
\end{abstract}

\section{Introduction} \label{sec: Introduction}

% Introduction and motivation
Glaucoma is a chronic optic neuropathy characterised by causing several visual field defects and structural changes in the optic nerve, such as a thinning of the retinal nerve fibre layer (RNFL) \cite{weinreb2004}. Nowadays, this degenerative disease is the leading cause of blindness worldwide and is expected to affect 111.8 million people in 2040 \cite{jonas2018}. The glaucoma diagnosis includes different expensive analysis (pachymetry, tonometry and visual field tests, among others) besides a subjective interpretation of expert ophthalmologists who often differ, especially in terms of early identification \cite{national2017}. Currently, imaging techniques based on fundus image and optical coherence tomography (OCT) have become a powerful tool to address the glaucoma diagnosis.

% Related work
\subsubsection{Related work.}
Timely treatment of glaucoma is essential to avoid the irreversible vision loss \cite{jonas2018}, so several computer-aided diagnosis systems and predictive algorithms focused on OCT and fundus images have been proposed in the literature to achieve early detection. Most of them were performed through traditional machine-learning (ML) techniques based on feature extraction and selection methods \cite{bizios2010, asaoka2017, kim2017}. All of them had in common the use of additional parameters relevant for glaucoma diagnoses, such as the intraocular pressure (IOP) and visual field (VF) tests, besides the OCT images. Unlike these works, we propose an innovative end-to-end system able to predict the glaucoma disease just from raw circumpapillary OCT images, without taking into account external expensive tests, like VF or IOP. We aim to elucidate the added value that this OCT samples around the retina optic nerve head (ONH) can provide for glaucoma detection. In recent years, the overwhelming irruption of the deep learning (DL) has replaced the traditional hand crafted-based methods, but most of the state-of-the-art studies used fundus images \cite{diaz2019, medeiros2019} or RNFL thickness probability maps extracted by combining fundus images and OCT B-scans \cite{muhammad2017, wang2019}. However, to the best of the authors' knowledge, we are the first that apply deep learning to evidence glaucoma just from raw circumpapillary OCT images.

\subsubsection{Contribution of this work.}
% Contribution of this work
In this paper, we propose a comparison between traditional and contemporary machine-learning models to analyse whether hand-driven approaches can outperform deep-learning algorithms for glaucoma detection, especially when addressing small databases. We hypothesise that the CNNs cannot replace the original way in which people can encode the information captured from a subjective point of view; in the same way that CNNs are able to identify hidden patterns that are not within reach of the human eye. For this reason, as the main novelty, we propose a hybrid model by fusing the features extracted from both ML and DL approaches to identify glaucoma.
\section{Material} \label{sec: Material}

The present work was carried out making use of a private database, coming from the Oftalvist Ophthalmic Clinic, which consists of 392 B-scans around the ONH of the retina. In particular, 194 samples from 97 patients were diagnosed by expert ophthalmologists as glaucomatous, whereas 198 circumpapillary images from 99 patients were associated with normal eyes. Note that each B-scan, of dimensions $M\times N=496\times768$ pixels, was acquired using the \textit{Heidelberg Spectrallis OCT} system, which allows obtaining an axial resolution of 4-5 $\mu$m.
\section{Methodology} \label{sec: Methodology}

\subsection{Data Partitioning} \label{subsec: Data_partitioning}
%In order to provide reliable results, we carried out a patient-based partitioning of the database to separate the training set from the independent test set. In particular, $\frac{4}{5}$ of the data (158 healthy and 156 glaucomatous OCT images from 79 and 78 patients, respectively) were used to train the predictive models, whereas $\frac{1}{5}$ (40 healthy and 38 glaucomatous B-scans from 20 and 19 patients) were employed to test them. Additionally, we applied an internal $k=5$-fold cross-validation technique intending to manage the overfitting and select the best parameters during the validation stage. Finally, we used the entire training set to build the final predictive models, which were assessed on the external test set.

In order to provide reliable results, we carried out a patient-based partitioning of the database to separate the training set from the independent test set. In particular, $\frac{4}{5}$ of the data (158 healthy and 156 glaucomatous OCT images) were used to train the predictive models, whereas $\frac{1}{5}$ (40 healthy and 38 glaucomatous B-scans) were employed to test them. Additionally, we applied an internal $k=5$-fold cross-validation technique intending to manage the overfitting and select the best parameters during the validation stage. Finally, we used the entire training set to build the final predictive models, which were assessed on the external set.

\subsection{Hand-driven learning methodology} \label{subsec: Hand_Driven_learning}

This approach consists of three main phases corresponding to the feature extraction, feature selection and classification through several traditional ML classifiers. Note that we used the RNFL and retina structures segmentation extracted by the Heidelberg Spectrallis system to perform this methodology.

\begin{figure}[h]
\begin{center}
\includegraphics[height=7cm, width=8.5cm]{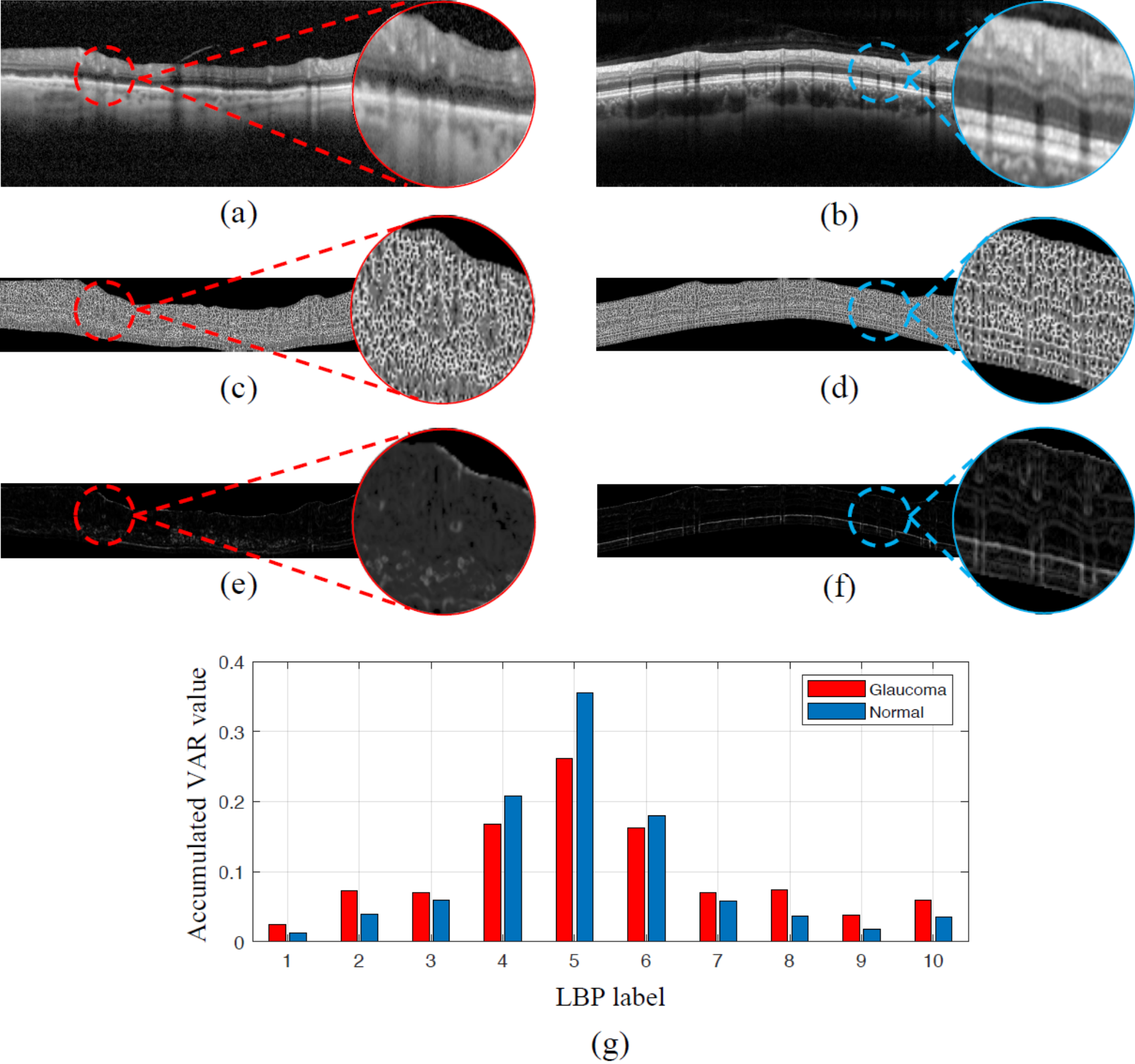}\\
\end{center}
\caption{(a-b) Examples of glaucomatous and normal samples. (c-d) LBP images. (e-f) VAR images. (g) 10-bin histograms of the LBPV operator from LBP and VAR images.}
\label{LBPs}
\end{figure}

\textbf{Feature extraction}.
For each circumpapillary OCT image $I_i$, being $i=\{1,2,3, ..., P\}$ and $P=392$ the number of B-scans, we combine, as a novelty, different variables related to four main descriptors: RNFL thickness, texture variables, fractal analysis and demographic data (age and gender). Regarding the RNFL thickness, we propose in this paper an innovative way of codifying the information, unlike \cite{asaoka2017, kim2017} where the authors employed the measures directly extracted from the hardware system. Let the vector $T_i=\{t_1, t_2, ..., t_j, ..., t_N\}$, where each $t_j \in T$ is the RNFL thickness calculated from the image $I_i$ in the position $j$, the proposed method is able to group the thickness values in a $h_{i,p}$ histogram vector, where $p=\{1,2,3,4\}$. For each $p$ value, $h_{i,p}$ vector allows quantifying the number of thickness $t_j$ whose distance is ranged between $D_p$ and $D_{p+1}$, being $D=[0,15,30,45,100]$ a vector of relevant distances, which were selected after studying the samples of the training set. Worth noting that minimum ($maxT_i$) and maximum ($minT_i$) RNFL thickness values were also calculated for this descriptor category.
Concerning the texture variables, grey-level co-occurrence matrix (GLCM) \cite{haralick1973} and local binary patterns (LBP) were applied for encoding the textural information contained in the retina region of each $I_i$. Variables such as contrast, correlation, energy, homogeneity, entropy, mean and standard deviation were calculated from a GLCM of dimensions $8\times 8$, using two offsets of [-2,0] and [-2,2] to measure the frequency in which a pixel with an intensity $x$ is adjacent to another with intensity $y$. Additionally, in order to recognise local texture information, we combined the uniformly invariant to rotation transforms $LBP_{P,R}^{riu2}$ operator, proposed in \cite{ojala2002}, with the rotational invariant local variance \textit{VAR} descriptor, to finally compute the LBP variance (LBPV) histogram, proposed in \cite{guo2010} (see Fig. \ref{LBPs}). It should be noted that similar texture descriptors have been previously considered for glaucoma detection from fundus images \cite{ali2014, kavya2017} but, to the best of the authors' knowledge, this is the first time that GLCM and LBP features are applied to circumpapillary OCT images.
Additionally, we analysed the fractal dimension in five directions (0º, 30º, 45º, 60º, 90º) via the Hurst Exponent \cite{hurst1965} computation to determine the presence of underlying trends in the complexity of the retinal region of each $I_i$. After the feature extraction phase, 75 hand-crafted variables per learning instance were taken into account to address the next stage. 

% %%%% FIGURE %%%%%%%% LBPs

\textbf{Feature selection}. An in-depth statistical analysis was carried out to select the most relevant variables in order to feed the proposed classifiers. Initially, a \textit{Kolmogorov-Smirnov test} was applied to determine the distribution of the variables. Then, \textit{Student{}'s t-tests} or \textit{Mann-Whitney U tests} were performed to analyse the discriminatory ability of each variable $v$ by comparing means or medians, respectively, depending on whether $v$ followed a normal distribution N(0,1) or not. The correlation coefficient was also calculated to obtain the independence grade between pairs of variables. Note that a level of significance $\alpha=0.05$ was defined for both hypothesis contrast to discard the non-relevant features, as well as the redundant information when \textit{p-value} $<\alpha$.

\textbf{Model Training}. Once the most relevant features were selected, we trained different ML classifiers such as support vector machine (SVM) and multi-layer perceptron (MLP), in line with the state-of-the-art studies \cite{bizios2010} and \cite{kim2017}. Several \textit{box constraints} and \textit{kernel scale} parameters for the SVM classifiers, and learning rates, loss functions, optimizers and network structures for the MLP network were considered during the internal cross-validation (ICV) stage. A considerable outperforming of the MLP classifier was achieved, with respect to the SVM, using the gradient descent adaptative optimizer with a learning rate of 0.001 and the binary cross-entropy as a loss function. Concerning the network structure, one hidden layer with 8 neurons reported the best model performance. Note that the proposed hand-driven learning approach is represented by the blue lines in the flowchart exposed in Fig. \ref{flowchart}.

\subsection{Deep-learning methodology} \label{subsec: Deep_learning}

%Similarly to the previous phase, an empirical exploration of several hyperparameters such as learning rates, batch sizes, loss functions and optimizers, was performed in order to build the best predictive deep-learning model during the ICV stage. Convolutional, pooling, dropout, batch normalisation and dense layers were also applied to specific experimental combinations in search of the best network architecture. Also, unlike the previously proposed ML approach, where the hand-crafted features were invariant to rotations and contrast transformations, here we considered the use of data augmentation techniques to alleviate the problem of insufficient data, by creating artificial samples via geometric and intensity modifications from the original images. In particular, the best performance was achieved using a CNN trained from scratch, which was composed of three blocks containing each one a ReLu-activated convolution layer followed by a $2\times 2$ max-pooling layer. Regarding the top model, a spatial squeeze was performed by a global max-pooling layer before the softmax layer with two neurons corresponding to the healthy and glaucoma classes. Adadelta optimizer with a learning rate of 0.005, squared hinge as a loss function and a batch size of 32 were selected as the best hyper-parameters during the ICV stage. It is noticeable that down-sampling $\times0.5$ of each OCT image was necessary to face the GPU memory constraints. The proposed deep-learning approach can be addressed by green lines in Fig. \ref{flowchart}. 

\begin{figure}[h]
\begin{center}
\includegraphics[height=2.5cm, width=8.5cm]{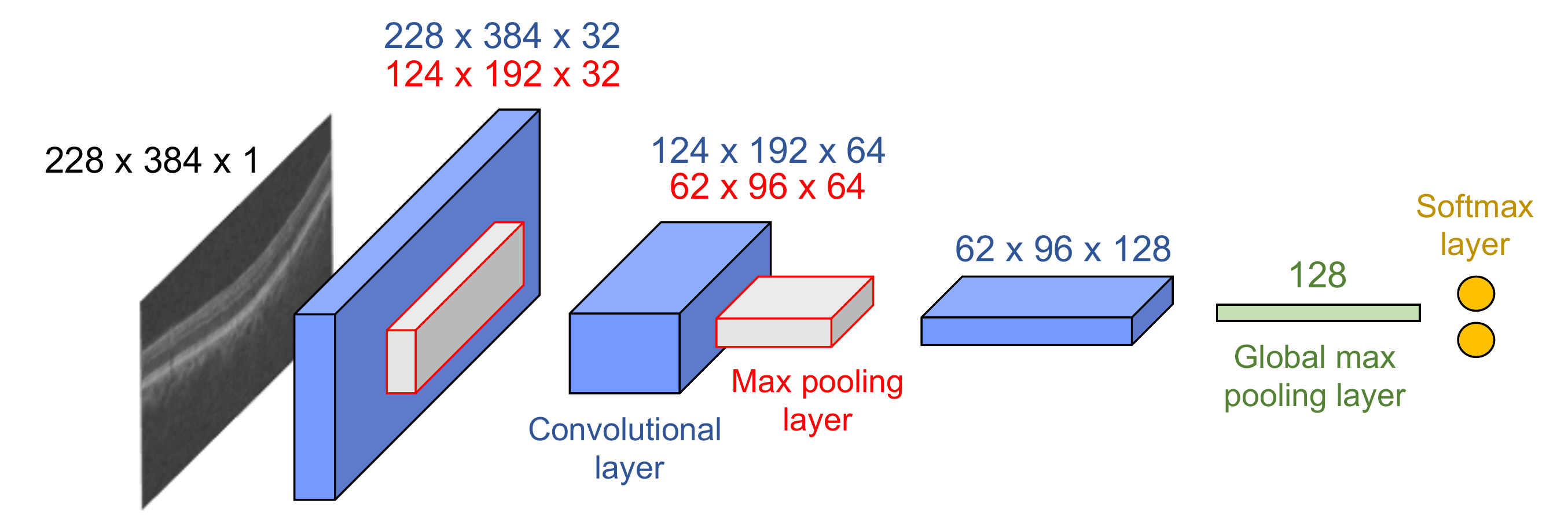}\\
\end{center}
\caption{Illustrative representation of the implemented CNN architecture.}
\label{CNN_architecture}
\end{figure}

Similarly to the previous phase, an empirical exploration of several hyperparameters was performed in order to build the best predictive deep-learning model during the ICV stage. Convolutional, pooling, dropout, batch normalisation and dense layers were also applied to specific experimental combinations in search of the best network architecture. Also, we considered the use of data augmentation techniques to alleviate the problem of insufficient data, by creating artificial samples via geometric and intensity modifications from the original images. The best performance was achieved by training the CNN exposed in Fig. \ref{CNN_architecture} and using Adadelta optimizer with a learning rate of 0.005, squared hinge as a loss function and a batch size of 32 during the ICV stage. It is noticeable that down-sampling $\times0.5$ of each OCT image was necessary to face the GPU memory constraints. The proposed deep-learning approach can be addressed by green lines in Fig. \ref{flowchart}. 

% \begin{table}[htbp]
% \caption{Proposed CNN architecture.}
% \label{CNN_architecture}
% %\renewcommand{\arraystretch}{1} % rows
% \setlength\tabcolsep{8 pt} % cols
% \small
% \begin{center}
% \begin{tabular}{ccc}
% \hline
% \textbf{Layer name} & \textbf{Output shape}                & \textbf{Filter size}            \\ \hline
% Input layer         & 248 x 384 x 1                        & N/A                             \\
% Conv1\_1            & 248 x 384 x 32                       & 3 x 3 x 32                      \\
% MaxPooling          & 124 x 192 x 32                       & 2 x 2 x 32                      \\
% Conv2\_1            & 124 x 192 x 64                       & 3 x 3 x 64                      \\
% MaxPooling          & 62 x 96 x 64                         & 2 x 2 x 64                      \\ 
% Conv3\_1            & 62 x 96 x 128                        & 3 x 3 x 128                     \\
% MaxGlobalPool       & 128                                  & N/A                             \\ 
% Dense (softmax)     & 2                                    & N/A                             \\ \hline
% \end{tabular}
% \end{center}
% \end{table}

% %%%% FIGURE %%%%%%%% Flowchart

\subsection{Hybrid approach} \label{subsec: HD+DL_approach}

\begin{figure}[h]
\begin{center}
\includegraphics[height=8cm, width=10cm]{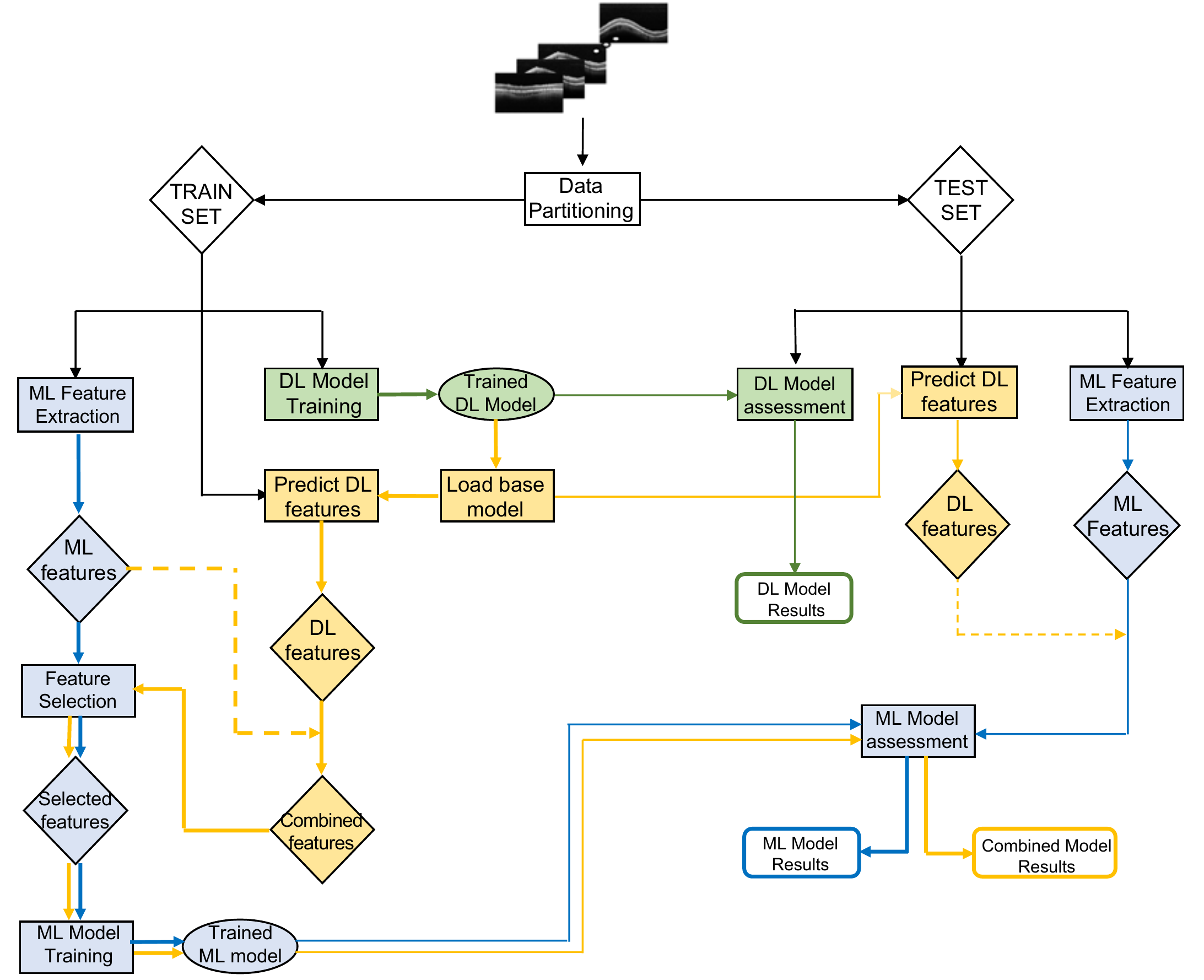}\\
\end{center}
\caption{Flowchart detailing the proposed ML, DL and hybrid approaches, in blue, green and yellow, respectively.}
\label{flowchart}
\end{figure}

As the main novelty in this paper, we propose a hybrid model to address the glaucoma detection taking into account both the hand-crafted and automatic-learning features. Our aim is to combine the original human point of view with the hidden potential enclosed in the CNNs. Specifically, we made use of the previously defined deep-learning base model as a feature extractor from each OCT image. Then, we fused the 75 ML and 128 DL extracted features to form the final feature vector from which we performed, in the same conditions, the feature selection and MLP training stages carried out in Section \ref{subsec: Hand_Driven_learning}. Finally, the three proposed models were assessed and compared using the test set, according to the flowchart exposed below. Note that the information relative to the hybrid approach can be interpreted by the yellow lines in Fig. \ref{flowchart}. 

% %%%% FIGURE %%%%%%%% Flowchart

\section{Results and discussion} \label{sec: Results}

\subsection{Feature selection results}

\begin{figure}[h]
\begin{center}
\begin{tabular}{cc}
\includegraphics[height=3cm, width=3.5cm]{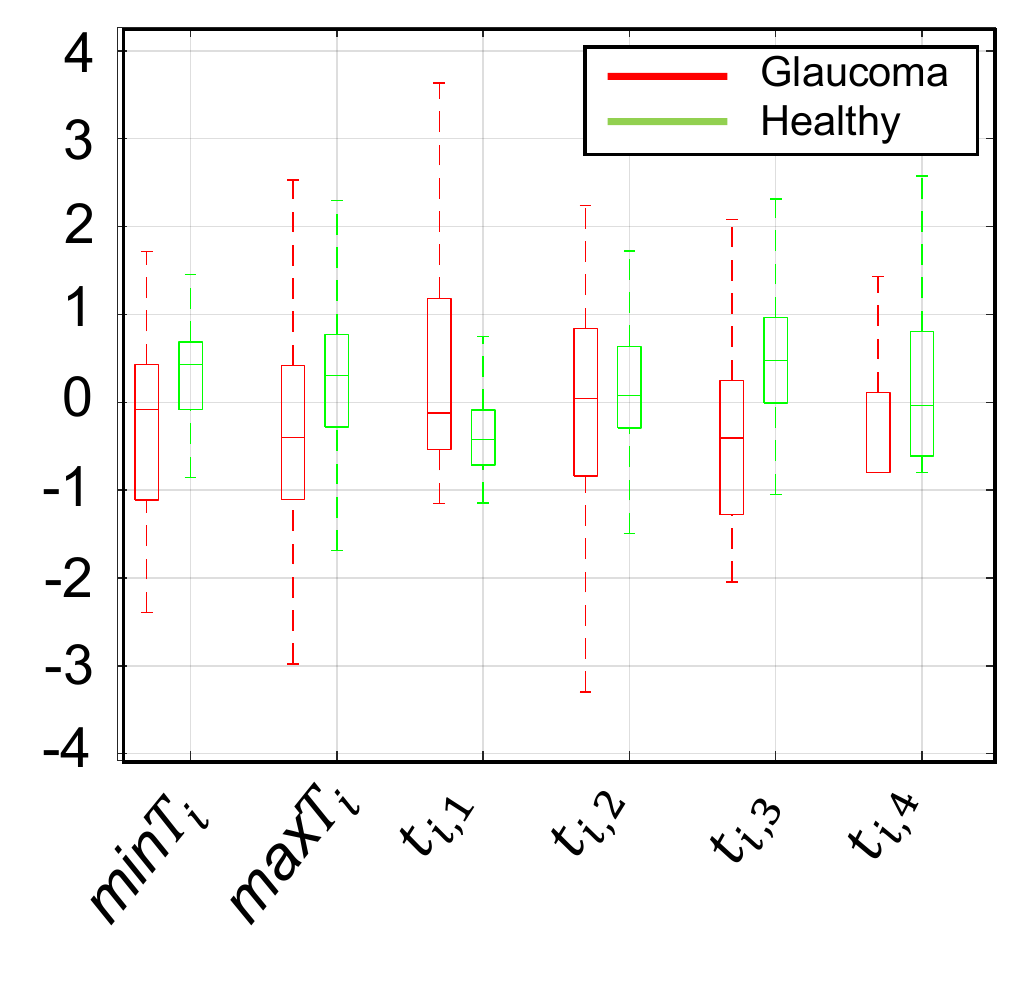} &
\includegraphics[height=3cm, width=3.5cm]{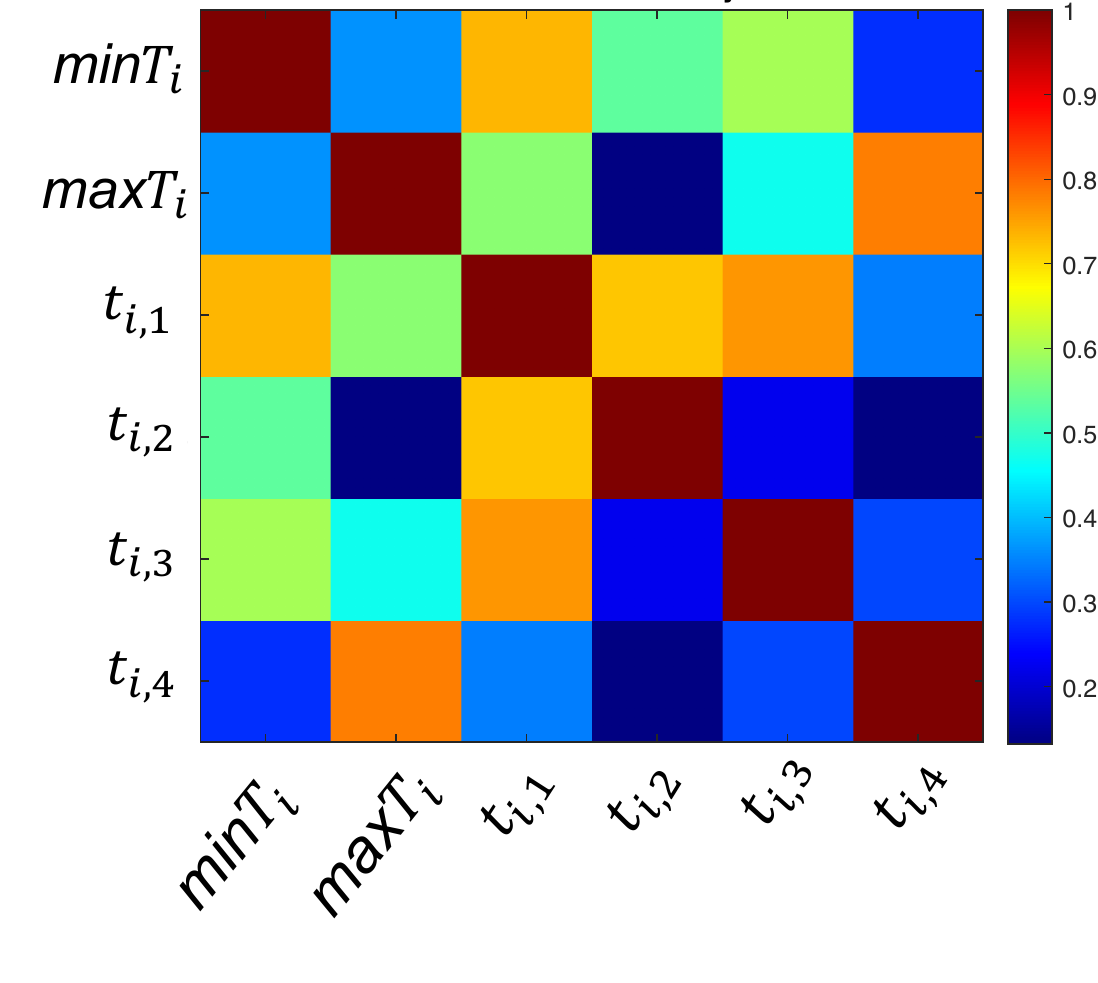} \\
(a) &
(b) \\
\end{tabular}
\end{center}
\caption{(a) Boxplot and (b) correlation matrix corresponding to the innovative six proposed RNFL thickness features.}
\label{statisticalResults}
\end{figure}

Regarding the hand-driven learning approach, 25 from a total of 75 features that composed a learning instance were selected after the statistical analysis. Otherwise, concerning the hybrid approach, 100 features from a total of 203 were reported as relevant variables to address the MLP training stage. Note that both ML and hybrid-final feature vectors included variables corresponding to the four kinds of descriptors used in this work. In addition, all the proposed features corresponding to the new RNFL thickness histogram-based method resulted statistically significant. A boxplot relative to these features is exposed in Fig. \ref{statisticalResults} to show the discriminatory ability of the proposed new descriptor. Besides, we also represent the correlation matrix of the same variables to evidence the independence level between them. 

%%%%%%%%%%%%% FIGURE %%%%%%%%%%%%%%

\subsection{Glaucoma prediction}

\textbf{Validation results}.
Classification results reached during the validation stage are detailed in Table \ref{ValidationResults} to objectively compare the proposed hand-driven learning (HDL), deep-learning (DL) and hybrid-learning methodologies. Different figures of merit, such as sensitivity (SN), specificity (SPC), F-score (FS), accuracy (ACC) and area under the ROC curve (AUC) are taken into account to assess the models providing reliable results. The findings are directly in line with the hypothesis postulated in Section \ref{sec: Introduction} since hand-driven learning approach has demonstrated surpassing deep-learning methods for small training sets, and the hybrid strategy clearly outperforms the rest, according to Table \ref{ValidationResults}.

\begin{table}[h]
\caption{Quantitative results reached during the ICV stage from all approaches.}
\label{ValidationResults}
\setlength\tabcolsep{6 pt}
%\small
\begin{center}
\begin{tabular}{cccc}
\hline
\multicolumn{1}{l}{}{} & \textbf{HDL approach}       & \textbf{DL approach}    & \textbf{Hybrid approach}\\
\hline
\textbf{SN}          & \textbf{0.802 $\pm$ 0.108}    & 0.747 $\pm$ 0.079       & 0.779 $\pm$ 0.110\\
\textbf{SPC}         & 0.807 $\pm$ 0.070             & 0.751 $\pm$ 0.080       & \textbf{0.912 $\pm$ 0.035}\\
\textbf{FS}          & 0.803 $\pm$ 0.072             & 0.748 $\pm$ 0.029       & \textbf{0.830 $\pm$ 0.074}\\
\textbf{ACC}         & 0.809 $\pm$ 0.058             & 0.748 $\pm$ 0.023       & \textbf{0.847 $\pm$ 0.048}\\
\textbf{AUC}         & 0.890 $\pm$ 0.056             & 0.823 $\pm$ 0.046       & \textbf{0.943 $\pm$ 0.018}\\
\hline
\end{tabular}
\end{center}
\end{table}

% %%%%%%%%%%%%% FIGURE %%%%%%%%%%%%%%
% \begin{figure}[h]
% \begin{center}
% \includegraphics[height=3cm, width=8cm]{Figures/ROCs_val.pdf} \\
% \end{center}
% \caption{ROC curves corresponding to the validation results.}
% \label{valROCs}
% \end{figure}

%The findings are directly in line with the postulated hypothesis in Section \ref{sec: Introduction}, since hand-driven learning methodology has demonstrated surpassing the deep-learning results when the training set is small, and the hybrid strategy clearly outperforms the rest, according to Fig. \ref{valROCs}. It should be noted that the hybrid approach reports considerable higher results in most of figures of merit, especially in specificity and PPV. However, HDL approach slightly stands out for the sensitivity and NPV measures, as it can be observed in Table \ref{ValidationResults}.

\textbf{Test results}.
In this section, we detail an external validation of the three proposed models using the independent test set, as it was previously explained in Fig. \ref{flowchart}. The classification results corresponding to the test set are exposed in Table \ref{testResults}. We can observe that, in line with the ICV stage, hand-driven learning provides a slight improvement regarding the deep-learning approach, which reaches values around 0.7 for all measures. Additionally, the hybrid methodology, characterised by the fusion of the features extracted from both ML and DL models, reports the most promising results for almost all figures of merit. It is important to note that an objective comparison with other state-of-the-art studies is not possible because all of them were performed on private databases or using another kind of input data, such as RNFL thickness probability maps or visual field tests.

\begin{table}[h]
\caption{Results comparison between the proposed models during the prediction stage.}
\label{testResults}
\setlength\tabcolsep{8 pt}
%\small
\begin{center}
\begin{tabular}{cccc}
\hline
\multicolumn{1}{l}{}{} & \textbf{HDL model}       & \textbf{DL model}    & \textbf{Hybrid model}\\
\hline
\textbf{SN}          & \textbf{0.7632}   & 0.6750       & 0.7368 \\
\textbf{SPC}         & 0.7500            & 0.6842      & \textbf{0.9000}\\
\textbf{FS}          & 0.7533            & 0.6835      & \textbf{0.8000}\\
\textbf{ACC}         & 0.7564            & 0.6795      & \textbf{0.8205}\\
\textbf{AUC}         & 0.8138            & 0.7480      & \textbf{0.8467}\\
\hline
\end{tabular}
\end{center}
\end{table}

\section{Conclusion} \label{sec: Conclusion}

In this work, three different learning methodologies have been proposed with the aim of elucidating that, under specific circumstances, hand-driven learning approaches can outperform deep-learning algorithms. The reported results evidenced that a combination of hand-crafted and data-learning strategies can improve the models' performance, especially for small databases. %5In future research lines, parameters such as VF tests and IOP could be included to build a robust computer-aided diagnosis system for glaucoma detection.

% ---- Bibliography ----

\bibliographystyle{splncs}

\bibliography{refs}

\begin{thebibliography}{10}

\bibitem{weinreb2004}
Weinreb, R.N., Khaw, P.T.:
\newblock Primary open-angle glaucoma.
\newblock The Lancet \textbf{363}(9422) (2004)  1711--1720

\bibitem{jonas2018}
Jonas, J.B., Aung, T., Bourne, R.R., Bron, A.M., Ritch, R., Panda-Jonas, S.:
\newblock Glaucoma--authors' reply.
\newblock The Lancet \textbf{391}(10122) (2018)  740

\bibitem{national2017}
National, G.A.U.:
\newblock Glaucoma: diagnosis and management.
\newblock (2017)

\bibitem{bizios2010}
Bizios, D., Heijl, A., Hougaard, J.L., Bengtsson, B.:
\newblock Machine learning classifiers for glaucoma diagnosis based on
  classification of retinal nerve fibre layer thickness parameters measured by
  stratus oct.
\newblock Acta ophthalmologica \textbf{88}(1) (2010)  44--52

\bibitem{asaoka2017}
Asaoka, R., Hirasawa, K., Iwase, A.e.a.:
\newblock Validating the usefulness of the “random forests” classifier to
  diagnose early glaucoma with optical coherence tomography.
\newblock American journal of ophthalmology \textbf{174} (2017)  95--103

\bibitem{kim2017}
Kim, S.J., Cho, K.J., Oh, S.:
\newblock Development of machine learning models for diagnosis of glaucoma.
\newblock PLoS One \textbf{12}(5) (2017)  e0177726

\bibitem{diaz2019}
Diaz-Pinto, A., Colomer, A., Naranjo, V., Morales, S., Xu, Y., Frangi, A.F.:
\newblock Retinal image synthesis and semi-supervised learning for glaucoma
  assessment.
\newblock IEEE transactions on medical imaging (2019)

\bibitem{medeiros2019}
Medeiros, F.A., Jammal, A.A., Thompson, A.C.:
\newblock From machine to machine: An oct-trained deep learning algorithm for
  objective quantification of glaucomatous damage in fundus photographs.
\newblock Ophthalmology \textbf{126}(4) (2019)  513--521

\bibitem{muhammad2017}
Muhammad, H., Fuchs, T.J., De~Cuir, N., De~Moraes, C.G.e.a.:
\newblock Hybrid deep learning on single wide-field optical coherence
  tomography scans accurately classifies glaucoma suspects.
\newblock Journal of glaucoma \textbf{26}(12) (2017)  1086

\bibitem{wang2019}
Wang, P., Shen, J., Chang, R., Moloney, M., Torres, M., Burkemper, B.e.a.:
\newblock Machine learning models for diagnosing glaucoma from retinal nerve
  fiber layer thickness maps.
\newblock Ophthalmology Glaucoma \textbf{2}(6) (2019)  422--428

\bibitem{haralick1973}
Haralick, R.M., Shanmugam, K., Dinstein, I.H.:
\newblock Textural features for image classification.
\newblock IEEE Transactions on systems, man, and cybernetics (6) (1973)
  610--621

\bibitem{ojala2002}
Ojala, T., Pietikainen, M., Maenpaa, T.:
\newblock Multiresolution gray-scale and rotation invariant texture
  classification with local binary patterns.
\newblock IEEE Transactions on pattern analysis and machine intelligence
  \textbf{24}(7) (2002)  971--987

\bibitem{guo2010}
Guo, Z., Zhang, L., Zhang, D.:
\newblock A completed modeling of local binary pattern operator for texture
  classification.
\newblock IEEE Transactions on Image Processing \textbf{19}(6) (2010)
  1657--1663

\bibitem{ali2014}
Ali, M.A., Hurtut, T., Faucon, T., Cheriet, F.:
\newblock Glaucoma detection based on local binary patterns in fundus
  photographs.
\newblock In: Medical Imaging 2014: Computer-Aided Diagnosis. Volume 9035.,
  International Society for Optics and Photonics (2014)  903531

\bibitem{kavya2017}
Kavya, N., Padmaja, K.:
\newblock Glaucoma detection using texture features extraction.
\newblock In: 2017 51st Asilomar Conference on Signals, Systems, and Computers,
  IEEE (2017)

\bibitem{hurst1965}
Hurst, H.E.:
\newblock Long term storage.
\newblock An experimental study (1965)

\end{thebibliography}

\end{document}